\documentclass[twocolumn,aps,pra,showpacs,superscriptaddress,floatfix]{revtex4}

\usepackage{color}
\usepackage{dcolumn}
\usepackage{graphicx}
\usepackage{epsf}
\usepackage{amsmath}
\usepackage{bm}
\usepackage{times}

\newcommand{\addrHD}{
Max--Planck--Institut f\"ur Kernphysik,
Saupfercheckweg 1, 69117 Heidelberg, Germany}

\newcommand{\addrNIST}{
National Institute of Standards and Technology,
Gaithersburg, Maryland 20899--8401}

\begin{document}

\sloppy 

\title{Two--Loop Bethe Logarithms for Higher Excited $S$ Levels}

\author{Ulrich D.~Jentschura}
\affiliation{\addrHD}
\affiliation{\addrNIST}

\begin{abstract}
Processes mediated by
two virtual low-energy photons contribute quite significantly 
to the energy of hydrogenic $S$ states. The corresponding level shift is 
of the order of $(\alpha/\pi)^2\,(Z\alpha)^6\,m_{\rm e}\,c^2$ 
and may be ascribed to a two-loop generalization of the 
Bethe logarithm.
For $1S$ and $2S$ states, the correction has recently been evaluated 
by Pachucki and Jentschura [Phys.~Rev.~Lett.~{\bf 91}, 113005 (2003)].
Here, we generalize the approach to higher excited $S$ states, 
which in contrast to the $1S$ and $2S$ states can decay to $P$
states via the electric-dipole ($E1$) channel. The more complex structure
of the excited-state wave functions and the necessity to subtract 
$P$-state poles lead to additional calculational problems.
In addition to the calculation
of the excited-state two-loop energy shift, we investigate 
the ambiguity in the energy level definition due to squared decay rates. 
\end{abstract}


\maketitle

%
%
\section{INTRODUCTION}
\label{intro} 

\begin{figure}[htb]%
\begin{center}\includegraphics[width=0.5\linewidth]{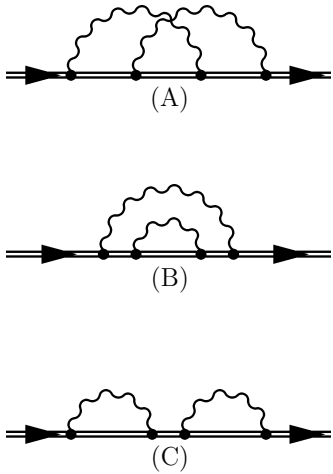}\end{center}%
\caption{\label{fig1}%
Two-photon processes may be interpreted in terms 
of Feynman diagrams. The double line denotes the 
bound-electron propagator. 
In the figure, we display the crossed-loop (A), the rainbow (B),
and the loop-after-loop (C) diagram.}%
\end{figure}

Both the experimental and 
the theoretical study of radiative corrections to bound-state energies
have been the subject of a continued endeavor over the last decades
(for topical reviews see~\cite{SaYe1990,MoPlSo1998,EiGrSh2001,Sh2002}).
Simple atomic systems like atomic hydrogen,
and heliumlike or lithiumlike systems, provide a testbed for 
our understanding of the fundamental interactions of light 
and matter, including the intricacies of the renormalization 
procedure and the complexities of the bound-state formalism.
One of the historically most problematic corrections for bound states
in hydrogenlike systems is the two-loop self-energy (2LSE) effect
(relevant Feynman diagrams in Fig.~\ref{fig1}), and this correction
will be the subject of the current paper.

%
%

Regarding self-energy calculations, 
two different approaches have been developed
for hydrogenlike systems: {\em (i)} the semianalytic approach, 
which is the so-called $Z\alpha$ expansion and in which 
the radiative corrections are expressed as a semianalytic
series expansion in the quantities $Z\alpha$
and $\ln[(Z\alpha)^{-2}]$, and {\em (ii)} the numerical approach, 
which avoids this expansion and leads to excellent accuracy for 
systems with a high nuclear charge number. Over the last couple 
of years, a number of calculations have been reported that profit from
recently developed 
numerical algorithms and an improved physical understanding of the 
problem at hand. These have led to numerical results even at 
low nuclear charge number~\cite{JeMoSo1999,JeMoSo2001pra,YeInSh2002}.

%
%
\begin{figure}[htb!]%
\begin{center}\includegraphics[width=0.9\linewidth]{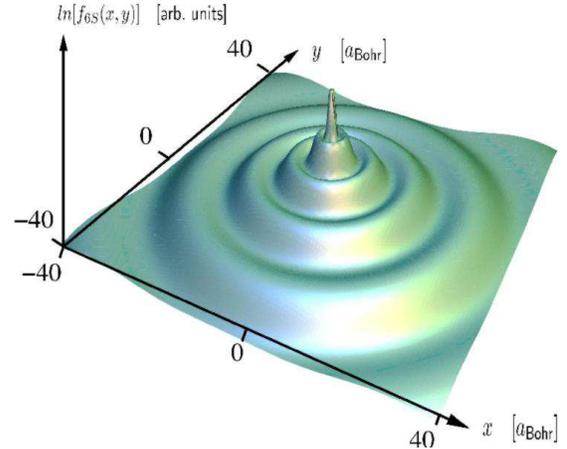}\end{center}%
\caption{\label{fig2}%
(color online).
Hydrogenic $S$ levels have the same (radial) symmetry properties
as the ground state. The wave function $\psi(r)$ of an $S$ level
therefore depends only on the radial coordinate $r \equiv \sqrt{x^2+y^2+z^2}$.
The function $f_{\rm 6S}(x,y) \equiv
\int^{\infty}_{-\infty} {\rm d}z \, \left| \psi_{\rm 6S}(x,y,z)\right|^2$
is positive definite 
and constitutes effectively an integrated projection of the $6S$ electron
probability density onto the $x$-$y$ plane. Indeed, we plot here the
natural logarithm of this function, which is
$\ln[f_{\rm 6S}(x,y)]$, as a function of $x \in [-40\,a_{\rm Bohr},
40\,a_{\rm Bohr}]$ and $y \in [-40\,a_{\rm Bohr}, 40\,a_{\rm Bohr}]$.
Here, $a_{\rm Bohr}$ denotes the Bohr radius 
$a_{\rm Bohr} = \hbar/(\alpha m c) = 0.529\,177\,2108(18) \times 
10^{-10}\,{\rm m}$~\cite{MoTaPriv2004}.}%
\end{figure}

Within approach {\em (i)},
a number of calculations have recently been reported which rely
on a separation of the energy scale(s) of the virtual photon(s) 
into high- and low-energy domains 
(see, e.g.,~\cite[Chap.~123]{BeLiPi1982},~\cite[Chap.~7]{ItZu1980}  
and~\cite{Pa1993}). This has recently
been generalized to two-loop corrections~\cite{Pa2001,JePa2002}. 
Also, there have been attempts 
to enhance our understanding of logarithmic 
corrections in higher orders 
of the $Z\alpha$ expansion by renormalization-group 
techniques~\cite{MaSt2000,Pi2002}. 

In the current paper, we discuss the evaluation of a specific
two-loop correction, which can quite naturally
be referred to as the two-loop generalization of the Bethe logarithm, 
for higher excited $S$ states (see also Fig.~\ref{fig2}). 
The calculation is carried out for the 
dominant nonlogarithmic contribution to the two-loop self-energy
shift of order $\alpha^2\,(Z\alpha)^6\,m\,c^2$,
where $m\,c^2$ is the rest energy of the electron
($m$ is the rest mass),
$\alpha$ is the fine-structure constant, and $Z$ is the 
nuclear charge number. 

The two-loop Bethe logarithm is formally of the same order 
of magnitude [$\alpha^2\,(Z\alpha)^6\,m\,c^2$] as 
a specific set of two-loop corrections which are mediated by squared decay
rates and whose physical interpretation has been shown to be limited
by the predictive power of the Gell--Mann--Low theorem on which 
bound-state calculations are usually based~\cite{JeEvKePa2002}.
For excited $nS$ states ($n \geq 3$), the problematic squared decay 
rates lead to an ambiguity which we assign to the two-loop Bethe logarithm 
as a further theoretical source of uncertainty.
Beyond the order of $\alpha^2\,(Z\alpha)^6\,m\,c^2$,
the definition of an atomic energy level becomes ambiguous, 
and the evaluation of radiative corrections
to energy levels has to be augmented by a more complete theory
of the line shape~\cite{Lo1952,LaSoPlSo2001,JeMo2002,JeEvKePa2002,
JeKePa2002,LaSoPlSo2002cjp,LaPrShBePlSo2002},
with the $1S$ state being the 
only true asymptotic state and 
therefore---in a strict sense---the only 
valid {\em in}- and {\em out}-state in the calculation of 
$S$-matrix type amplitudes~\cite{JeEvKePa2002}. Further
interesting thoughts on questions
related to line shape profiles can be found in~\cite{Mo1978,HiMo1980}.
It had also been pointed out in Sec.~VI of~\cite{JeSoMo1997} that
the asymmetry of the natural line shape has to be considered at the 
order $\alpha^8$.

It is tempting to ask how
one may intuitively understand the slow convergence of the
$Z\alpha$ expansion of the two-loop energy shift.
As pointed out in~\cite{EiGrSh2001},
terms of different order in the $Z\alpha$ expansion
have rather distinct physical origins.
In the order $\alpha^2\,(Z\alpha)^4\,m\,c^2$,
there are two corrections, both arising from
hard (high-energy) virtual photons. These correspond,
respectively, to the infrared {\em convergent}
slope of the two-loop electron Dirac form factor and
to a two-loop anomalous magnetic moment correction.
The term of order $\alpha^2\,(Z\alpha)^5\,m\,c^2$
may also be computed without any consideration
of low-energy virtual 
quanta~\cite{Pa1993,Pa1994prl,EiGrPe1994,EiSh1995,EiGrSh2001}.
The terms of order $\alpha^2\,(Z\alpha)^6\,m\,c^2$
are not the leading terms arising from the two-loop 
self-energy shifts. Logarithmic correction terms of order 
$\alpha^2\,(Z\alpha)^6\,\ln^i[(Z\alpha)^{-2}]\,m\,c^2$ 
($i=1,2,3$) have been considered in~\cite{Ka1996,Pa2001}. It is only at the 
order of $\alpha^2\,(Z\alpha)^6\,m\,c^2$ that the low-energy
virtual photons begin to contribute to the hydrogenic energy shift(s) of 
$S$ states. They do so quite significantly, enhanced by the 
triple logarithm ($i=3$) and a surprisingly large coefficient
of the single logarithm ($i=1$), as shown in~\cite{Pa2001}.
It is therefore evident that we need to gain an understanding
of all logarithmic and nonlogarithmic terms ($i=0,1,2,3$) of order
$\alpha^2\,(Z\alpha)^6\,\ln^i[(Z\alpha)^{-2}]\,m\,c^2$ before
any reliable prediction for the two-loop 
bound-state correction can be made even at low nuclear charge number.

An interesting observation can be made
based on the fact that the 
imaginary part of the (nonrelativistic) one-loop self-energy
gives the leading-order contribution to the $E1$ one-photon decay
width of excited states~\cite{BaSu1978}.
Analogously, it is precisely the imaginary part of the 
nonrelativistic two-loop self-energy which corresponds to 
the two-photon decay rate of the $2S$ state.
The $2S$ two-photon decay rate is of the 
order of $\alpha^2\,(Z\alpha)^6\,m\,c^2$
(see, e.g.,~\cite{ShBr1959,Kl1969,CrTaSaCh1986,FlScMi1988}).
From a nonrelativistic point of view,
the scaling $\alpha^2\,(Z\alpha)^6\,m\,c^2$ 
can be seen as some kind of ``natural''
order for the two-loop effect.
It is the first order in which logarithms of $Z\alpha$
appear and the first order in which a matching of
low- and high-energy contributions is required.
This is also reflected in the properties of the 
two-photon decay.

This article is organized as follows.
In Sec.~\ref{knowncoeff}, we review the status of known
two-loop self-energy corrections.
The formulation of the problem in nonrelativistic
quantum electrodynamics (NRQED) and 
calculation is discussed in Sec.~\ref{2Lnrqed}.
Squared decay rates are the subject of Sec.~\ref{ambiguity},
and further contributions to the 
self-energy in the order of $\alpha^2\,(Z\alpha)^6\,m\,c^2$ are
discussed in Sec.~\ref{further}.
Finally, conclusions are drawn in Sec.~\ref{conclu}.

%
%
\section{KNOWN TWO--LOOP SELF--ENERGY COEFFICIENTS}
\label{knowncoeff}

We work in natural units ($\hbar = c = \epsilon_0 = 1$),
as is customary in QED bound-state calculations.
The (real part of the) level shift of a hydrogenic state
due to the two-loop self-energy reads
\begin{eqnarray}
\label{DefESE2L}
\Delta E^{\rm (2L)}_{\mathrm{SE}} &=& \left(\frac{\alpha}{\pi}\right)^2 \,
\frac{(Z\alpha)^4\,m_{\rm e}}{n^3} \, H(Z\alpha)\,.
\end{eqnarray}
For the two-loop self-energy 
(2LSE) diagrams (see Fig.~\ref{fig1}), the first terms
of the 
semianalytic expansion in powers of $Z\alpha$
and $\ln[(Z\alpha)^{-2}]$ read
\begin{eqnarray}
\label{defH}
\lefteqn{H(Z\alpha) = B^{\rm (2LSE)}_{40} +
(Z\alpha)\, B^{\rm (2LSE)}_{50} }
\nonumber\\[0.5ex]
& & + (Z\alpha)^2 \,
\left\{ B^{\rm (2LSE)}_{63} \, \ln^3(Z\alpha)^{-2} 
+ B^{\rm (2LSE)}_{62} \, \ln^2(Z\alpha)^{-2} \right.
\nonumber\\[0.5ex]
& &
\left.
+ B^{\rm (2LSE)}_{61} \, \ln(Z\alpha)^{-2} + B^{\rm (2LSE)}_{60} \right\} \,.
\end{eqnarray}
The function $H(Z\alpha)$ is dimensionless.
We ignore unknown higher-order terms
in the $Z\alpha$ expansion 
and focus on a specific numerically large contribution
to $B^{\rm (2LSE)}_{60}$ given by the two-loop Bethe 
logarithm. We also keep the upper index (2LSE) in order to distinguish the 
two-loop self-energy contributions to the analytic coefficients from the 
self-energy vacuum-polarization (SEVP) effects~\cite{Pa2001,Je2003plb}
and the vacuum-polarization insertion into the virtual photon line 
in the one-loop self-energy (SVPE). By contrast,
the sum of these effects carries no upper index, 
according to a convention
adopted previously in~\cite{Pa2001,Je2003plb}. It has been mentioned
earlier that $B_{40}$ and $B_{50}$ are purely relativistic effects
mediated by hard virtual photons.
The coefficient
$B_{40}$ in Eq.~(\ref{defH}) 
involves a Dirac and a Pauli form factor correction
and reads~\cite{TwoLoop}
\begin{equation}
\label{B40}
B^{\rm (2LSE)}_{40}(nS) =
- \frac{163}{72} - \frac{85}{36}\,\zeta(2) +
9\,\ln(2)\,\zeta(2) - \frac94 \, \zeta(3) \,;
\end{equation}
the numerical value is $1.409\,244$.
The first relativistic correction $B^{\rm (2LSE)}_{50}(nS)$
is known to have a rather large value~\cite{Pa1994prl},
\begin{equation}
\label{B50}
B^{\rm (2LSE)}_{50}(nS) = - 24.2668(31)\,. 
\end{equation}
The triple logarithm in the sixth order of $Z\alpha$ reads,
\begin{equation}
\label{B63}
B_{63}(nS) = B^{\rm (2LSE)}_{63}(nS) = -\frac{8}{27}\,.
\end{equation}
It has meanwhile been clarified~\cite{Ye2000,Ye2001,JeNa2002,YeInSh2003}
that the total value of this
coefficient is the result of subtle cancellations among 
the different diagrams displayed in Fig.~\ref{fig1}. 
The double logarithm for $nS$ is given by
\begin{eqnarray}
\label{B62}
B^{\rm (2LSE)}_{62}(1S) &=& \frac{16}{27} - \frac{16}{9}\, \ln(2)
= -0.639\,669\,, \\[0.5ex]
B^{\rm (2LSE)}_{62}(nS) &=& B^{\rm (2LSE)}_{62}(1S)
\nonumber\\[2ex]
& & \!\!\!\!\!\!\!\!\!\!\!\!\!\!\!\!\!\!\!\!\!\!\!\!\!\!\!\!\!\!\!\!
+ \frac{16}{9} \left( \frac34 + \frac{1}{4 n^2} -
\frac1n - \ln(n) \! + \! \Psi(n) + C\right),
\end{eqnarray}
where $\Psi$ denotes the logarithmic derivative of the gamma function,
and $C = 0.577216\dots$ is Euler's constant.

The result for $B_{61}$, restricted to the 
two-loop diagrams in Fig.~\ref{fig1}, 
reads~\cite{Pa2001,Je2003plb}
\begin{eqnarray}
\label{B61}
B^{\rm (2LSE)}_{61}(1S) 
&=& \frac{5\,221}{1\,296}
+ \frac{875}{72}\,\zeta(2) + \frac92\, \,\zeta(2)\,\ln 2
\nonumber\\[0.5ex]
& & - \frac98 \, \zeta(3)
- \frac{152}{27}\,\ln 2
+ \frac{40}{9} \, \ln^2 2 
\nonumber\\[0.5ex]
& & + \frac43 \, N(1S)
\nonumber\\[0.5ex]
&=& 49.838\,317 \,, \\[0.5ex]
\label{ndep}
B^{\rm (2LSE)}_{61}(nS) &=& B^{\rm (2LSE)}_{61}(1S)
+ \frac43 \, \left[ N(nS) - N(1S)\right]
\nonumber\\[0.5ex]
& & + \! \left( \frac{80}{27} - \frac{32}{9} \, \ln 2\right)\,
\left(\frac34 + \frac{1}{4 n^2} \right.
\nonumber\\[0.5ex]
& & \left. - \frac1n - \ln(n) \! + \! \Psi(n)
\! + \! C\right)\,.
\end{eqnarray}
We correct here a calculational error in Eq. (7a) of Ref.~\cite{Je2003plb} 
where a
result of $49.731651$ had been given for $B^{\rm (2LSE)}_{61}(1S)$.
However, even with this correction, the result for
$B^{\rm (2LSE)}_{61}(1S)$ given in
Eq.~(\ref{B61}) is incomplete because it lacks contributions from
two-Coulomb-vertex diagrams. These diagrams give rise to an effective
interaction proportional to $\bm{E}^2$ in the NRQED Hamiltonian and will
be discussed in~\cite{PaJeMaBo2004}. The additional contribution
to $B^{\rm (2LSE)}_{61}(1S)$ does not affect the $n$-dependence of this
coefficient as indicated in Eq.~(\ref{ndep}),
nor does it affect the calculation of
the two-loop Bethe logarithms presented in this article.

Numerical values of $N(nS)$ are given in~\cite[Eq.~(12)]{Je2003jpa}
for $n= 1,\dots,8$:
\begin{subequations}
\label{resultsnS}
\begin{eqnarray}
N(1S) &=& 17.855\,672(1)\,, \\[1ex]
N(2S) &=& 12.032\,209(1)\,, \\[1ex]
N({3S}) &=& 10.449\,810(1)\,, \\[1ex]
N(4S) &=& 9.722\,413(1)\,, \\[1ex]
N(5S) &=& 9.304\,114(1)\,, \\[1ex]
N(6S) &=& 9.031\,832(1)\,, \\[1ex]
N(7S) &=& 8.840\,123(1)\,, \\[1ex]
N(8S) &=& 8.697\,639(1)\,. 
\end{eqnarray}
\end{subequations}

%
%
\section{TWO--LOOP PROBLEM IN NRQED}
\label{2Lnrqed}

Historically,
one of the first two-photon problems to be tackled theoretically
in atomic physics has been the two-photon decay of the metastable 
$2S$ level which was treated in ~\cite{GO1929,GM1931}. 
It is this decay channel which limits the lifetime 
of the $2S$ hydrogenic state. We have~\cite{ShBr1959}
\begin{equation}
\tau^{-1} = \Gamma = 8.229 \, Z^6\, {\rm s}^{-1} = 1.310 \, Z^6\, {\rm Hz}\,.
\end{equation}
The numerical prefactors of the width is different
when expressed in inverse seconds and alternatively in Hz.
The following remarks are meant to clarify this situation
as well as the entries in Table~\ref{table2} below.
In order to obtain the width in Hz, one should interpret the
imaginary part of the self-energy~\cite{BaSu1978} as $\Gamma/2$,
and do the same conversion as for the real part of the
energy, i.e. divide by $h$, not $\hbar$.
This gives the width in Hz. The unit Hz corresponds
to cycles per second.

In order to obtain the lifetime in inverse seconds, 
which is radians per second,
one has to multiply the previous result by
a factor of $2 \pi$, a result which may
alternatively be obtained by dividing $\Gamma$---i.e.~the 
imaginary part of the energy, by $\hbar$, not $h$. 
The general paradigm is that in order to evaluate an energy in 
units of Hz,
one should use the relation $E = h \, \nu$, whereas for a
conversion of an imaginary part of an energy to the inverse 
lifetime, one should use $\Gamma = \hbar \, \tau^{-1}$.
As calculated in Refs.~\cite{ShBr1959,Kl1969}, 
the width of the metastable 
$2S$ state in atomic hydrogenlike systems is $8.229\,Z^6\,{\rm s}^{-1}$ 
(inverse seconds). At $Z=1$,
this is equivalent to the ``famous'' value of
$1.3\,{\rm Hz}$ which is nowadays most 
frequently quoted in the literature.
The lifetime of a hydrogenic $2S$ level is thus
$0.1215\,Z^{-6}\,{\rm s}$.
This latter fact has been verified experimentally for 
ionized helium~\cite{Pr1972,KoClNo1972,HiClNo1978}.

We now briefly recall the expression for the 
two-photon decay involving two emitted photons with 
polarization vectors $\bm{\varepsilon}_1$ and $\bm{\varepsilon}_2$, 
in a two-photon transition from an initial 
state $|\phi_i\rangle$ to a final state $|\phi_f\rangle$.
The two-photon decay width $\Gamma$ is given by
[see, for example,~Eq.~(3) of Ref.~\cite{ShBr1959}]
\begin{eqnarray}
\label{tpd1}
\Gamma &=& \frac{4}{27}\,\frac{\alpha^2}{\pi}\,
\int\limits^{\omega_{\rm max}}_0 
{\rm d}\omega_1 \, \omega^3_1\,\omega^3_2 \,
\nonumber\\[1ex]
& & \left| \left< \phi_f \left| 
x^i \, \frac{1}{H - E + \omega_2} \, x^i \right| \phi_i \right> \right.
\nonumber\\[1ex]
& & \left.
+ \left< \phi_f \left| 
x^i \, \frac{1}{H - E + \omega_1} \,
x^i \right| \phi_i \right> 
\right|^2\,,
\end{eqnarray}
where $\omega_2 = \omega_{\rm max} - \omega_1$
and $\omega_{\rm max} = E - E'$ is the maximum energy 
that any of the two photons may have.
The Einstein summation convention is used throughout this article.
Note the identity~\cite{BaFoQu1977,Ko1978prl}
\begin{eqnarray}
\label{tpd2}
\lefteqn{\left< \phi_f \left|
\frac{p^i}{m} \, \frac{1}{H - E + \omega_1} \,
\frac{p^i}{m} \right| \phi_i \right> }
\nonumber\\[1ex]
& & 
+ \left< \phi_f \left| 
\frac{p^i}{m} \, \frac{1}{H - E + \omega_2} \,
\frac{p^i}{m} \right| \phi_i \right> 
\nonumber\\[2ex]
&=&
-\omega_1\,\omega_2\,m^2\,
\left\{ \left< \phi_f \left| 
x^i \, \frac{1}{H - E + \omega_1} \,
x^i \right| \phi_i \right> \right.
\nonumber\\[1ex]
& & \left.
+ \left< \phi_f \left| 
x^i \, \frac{1}{H - E + \omega_2} \,
x^i \right| \phi_i \right> \right\}\,,
\end{eqnarray}
which is valid at exact resonance $\omega_1 + \omega_2 = E_i - E_f$.
This identity permits a reformulation of the problem in the 
velocity-gauge as opposed to the length-gauge form.

In a number of cases, the formulation of a quantum electrodynamic 
bound-state problem may be simplified drastically when
employing the concepts of an effective low-energy
field theory known as nonrelativistic quantum electrodynamics~\cite{CaLe1986}. 
The basic idea consists in a correspondence between 
fully relativistic quantum electrodynamics and effective low-energy
couplings between the electron and radiation field, which still
lead to ultraviolet divergent expressions. However, the 
ultraviolet divergences may be matched against effective 
high-energy operators, which leads to a cancellation of the 
cut-off parameters. Within the context of the 
{\em one}-loop self-energy problem,
a specialized approach has been discussed 
in~\cite{Pa1993,JePa1996,JeSoMo1997,Je2003dipl}.
The formulation of the {\em two}-loop self-energy problem 
within the context of nonrelativistic quantum 
electrodynamics (NRQED) has been discussed in~\cite{Pa2001}.
We denote by $p^j$ the Cartesian components of the momentum 
operator $\bm{p} = -{\rm i}\,\bm{\nabla}$.
The expression for the two-loop self-energy shift 
reads~\cite{Pa2001,remarkPRA2002}
\begin{widetext}
\begin{eqnarray}
\label{NRQED}
\lefteqn{
\Delta E_{\rm NRQED} = 
- \left( \frac{2 \, \alpha}{3 \,\pi\,m^2} \right)^2 \,
\int_0^{\epsilon_1} {\rm d}\omega_1 \, \omega_1 \,
\int_0^{\epsilon_2} {\rm d}\omega_2 \, \omega_2 \, 
\left\{  
\left< p^i \, \frac{1}{H - E + \omega_1} \, p^j \, 
\frac{1}{H - E + \omega_1 + \omega_2} \, p^i \, 
\frac{1}{H - E + \omega_2} \, p^j \right> \right.} \nonumber\\[1ex]
& & + \frac{1}{2} \,
\left< p^i \, \frac{1}{H - E + \omega_1} \, p^j \,
\frac{1}{H - E + \omega_1 + \omega_2} \, p^j \, 
\frac{1}{H - E + \omega_1} \, p^i \right> \nonumber\\[1ex]
& & + \frac{1}{2} \,
\left< p^i \, \frac{1}{H - E + \omega_2} \, p^j \,
\frac{1}{H - E + \omega_1 + \omega_2} \, p^j \, 
\frac{1}{H - E + \omega_2} \, p^i \right>
\nonumber\\[1ex]  
& & + 
\left< p^i \, \frac{1}{H - E + \omega_1} \, p^i \, 
\left( \frac{1}{H - E} \right)' \, p^j \, 
\frac{1}{H - E + \omega_2} \, p^i \right> 
\nonumber\\[1ex]  
& & - \frac{1}{2} \,
\left< p^i \, \frac{1}{H - E + \omega_1} \, p^i \right> \,
\left< p^j \, \left( \frac{1}{H - E + \omega_2} \right)^2 \, p^i \right>
- \frac{1}{2} \,
\left< p^i \, \frac{1}{H - E + \omega_2} \, p^i \right> \,
\left< p^j \, \left( \frac{1}{H - E + \omega_1} \right)^2 \, p^i \right>
\nonumber\\[1ex]
& & \left. - m \,
\left< p^i \, \frac{1}{H - E + \omega_1} \, 
\frac{1}{H - E + \omega_2} \, p^i \right>
- \frac{m}{\omega_1 + \omega_2} \,
\left< p^i \, \frac{1}{H - E + \omega_2} \, p^i \right>
- \frac{m}{\omega_1 + \omega_2} \,
\left< p^i \, \frac{1}{H - E + \omega_1} \, p^i \right>
\right\}.
\end{eqnarray}
\end{widetext}
All of the matrix elements are evaluated on the reference state
$|\phi\rangle$, for which the nonrelativistic Schr\"{o}dinger wave
function is employed. The Schr\"{o}dinger Hamiltonian is denoted by $H$,
and $E = -(Z\alpha)^2 \, m/(2 n^2)$ is the Schr\"{o}dinger energy
of the reference state ($n$ is the principal quantum number).

Expressions~(\ref{tpd1}) and~(\ref{tpd2}) now follow in a natural
way as the imaginary part generated by the sum of the 
first three terms in curly brackets in
Eq.~(\ref{NRQED}). Specifically, the poles are generated
upon $\omega_2$-integration by
the propagator 
\begin{equation}
\frac{1}{H - E + \omega_1 + \omega_2} = 
\sum_{j} \frac{| j \rangle \, \langle j |}
  {E_j - E  + \omega_1 + \omega_2}
\end{equation} 
when $E - E_j = \omega_1 + \omega_2$,
which is just the energy conservation 
condition for two-photon decay.
Of course, other terms in Eq.~(\ref{NRQED}), not just the 
first three in curly brackets, may also generate 
imaginary parts (especially if the reference
state is an excited state, and one-photon decay is possible).
The corresponding pole terms must be dealt with in 
a principal-value prescription if we are
interested only in the real part of the energy shift. 
For $P$ states and higher excited $S$ states,
the remaining imaginary parts find a natural interpretation
as radiative correction
to the one-photon decay width~\cite{SaPaCh2004}.

From here on we scale the photon frequencies $\omega_{1,2}$ by
\begin{subequations}
\label{scaling}
\begin{equation}
\label{scaling1}
\omega_k \to \omega'_k \equiv \frac{\omega_k}{(Z\alpha)^2 \, m}
\,,\quad k = 1,2\,.
\end{equation}
This scaling, which is convenient for our numerical calculations,
(almost) corresponds to a transition to
atomic units [but with $(Z\alpha)^2\,m = 1$ instead of
a unit Rydberg constant].
The momentum operator is scaled as
\begin{equation}
\label{scaling2}
\bm{p} \to  \bm{p}' \equiv \frac{\bm{p}}{Z\alpha\,m}
\end{equation}
and becomes a dimensionless quantity. The Schr\"odinger Hamiltonian
is scaled as 
\begin{equation}
\label{scaling3}
H \to H' \equiv \frac{H}{(Z\alpha)^2 \, m}\,.
\end{equation}
The binding energy of the reference state
receives a scaling as 
\begin{equation}
\label{scaling4}
E \to E' \equiv \frac{E}{(Z\alpha)^2 \, m} = -\frac{1}{2 n^2} 
\end{equation}
and is from now on also a dimensionless quantity
($n$ is the principal quantum number). 
The scaled, dimensionless radial coordinate is 
obtained as 
\begin{equation}
\label{scaling5}
r \to r' \equiv Z\alpha\,m\,r \,.
\end{equation}
The scaled Green function
\begin{equation}
\label{defG}
G'(\omega') =  \frac{1}{E' - H' - \omega'}
\end{equation}
is also dimensionless. Finally, the quantity
\begin{equation}
G'_{\rm red}(0) = 
\sum_{|j\rangle \neq |\phi\rangle} 
\frac{| j \rangle \, \langle j|}{E' - E'_j} 
\end{equation}
\end{subequations}
is the reduced Green function where the reference state 
$|\phi\rangle$ is excluded from the sum over intermediate 
states. The (scaled dimensionless) Schr\"odinger
Hamiltonian is then given as $H'=\bm{p}'^2/2-1/r'$.
Scaled quantities will be used from here on until the end 
of the current Section~\ref{2Lnrqed}, 
and we will denote the scaled, dimensionless
quantities by primes, for absolute clarity of notation.
(Note that in
Ref.~\cite{PaJe2003}, the corresponding scaled quantities were denoted
by the same symbol as the dimensionful quantities.)
As indicated in~\cite[Eq.~(5)]{PaJe2003},
the expression (\ref{NRQED}) can be rewritten in
terms of the scaled quantities as
\begin{equation}
\label{NRQED2}
\Delta E_{\rm NRQED} = 
\frac49\,
\left(\frac{\alpha}{\pi}\right)^2\, (Z\alpha)^6 \,m\,
\int {\rm d}\omega'_1 \int {\rm d}\omega'_2 \,
f(\omega'_1, \omega'_2) \,,
\end{equation}
where the (dimensionless) function $f(\omega'_1, \omega'_2)$ is defined as
[see also Eq.~(\ref{NRQED})]
\begin{widetext}
\begin{eqnarray}
& & f(\omega'_1, \omega'_2) = 
\omega'_1 \, \omega'_2\, 
\left[ \left< p'^i \, G'(\omega'_1) \, p'^j \,
G'(\omega'_1 + \omega'_2) \, p'^i \,
G'(\omega'_2) \, p'^j \right> 
+ \frac{1}{2} \,
\left< p'^i \, G'(\omega'_1) \, p'^j \,
G'(\omega'_1 + \omega'_2) \, p'^j \,
G'(\omega'_1) \, p'^i \right> \right. \nonumber\\[1ex]
& & + \frac{1}{2} \,
\left< p'^i \, G'(\omega'_2) \, p'^j \,
G'(\omega'_1 + \omega'_2) \, p'^j \,
G'(\omega'_2) \, p'^i \right> +
\left< p'^i \, G'(\omega'_1) \, p'^i \,
G'_{\rm red}(0) \, p'^j \,
G'(\omega'_2) \, p'^i \right>
\nonumber\\[1ex]
& & - \frac{1}{2} \,
\left< p'^i \, G'(\omega'_1) \, p'^i \right> \,
\left< p'^j \, G'^2(\omega'_2) \, p'^i \right>
- \frac{1}{2} \,
\left< p'^i \, G'(\omega'_2) \, p'^i \right> \,
\left< p'^j \, G'^2(\omega'_1) \, p'^i \right>
\nonumber\\[1ex]
& & \left. + 
\left< p'^i \, G'(\omega'_1) \,
G'(\omega'_2) \, p'^i \right>
- \frac{1}{\omega'_1 + \omega'_2} \,
\left< p'^i \, G'(\omega'_2) \, p'^i \right>
- \frac{1}{\omega'_1 + \omega'_2} \,
\left< p'^i \, G'(\omega'_1) \, p'^i \right> \right] \,.
\end{eqnarray}
\end{widetext}
In~\cite{PaJe2003}, the corresponding Equation~(5) has a 
typographical error: the term $m \, \left< p'^i \, G'(\omega'_1) \,
G'(\omega'_2) \, p'^i \right>$ should have a plus instead of a 
minus sign (seventh term in the square brackets). 
In particular, the scaling (\ref{scaling})
leads to a disappearance of the powers of $Z\alpha$ when considering
the expression $\int {\rm d}\omega'_1 \, \omega'_1 \,
\int {\rm d}\omega'_2 \, \omega'_2 \,                      
f(\omega'_1, \omega'_2)$.

First, we fix $\omega'_1$ and integrate over $\omega'_2$.
The subtraction procedure is as follows.
We need to subtract the contribution from the following terms
that lead to divergent expressions as $\omega'_2 \to \infty$. 
We therefore expand $f(\omega'_1, \omega'_2)$ for large 
$\omega'_2$ at fixed $\omega'_1$. The asymptotics read~\cite{PaJe2003}
\begin{eqnarray}
\label{largew2}
f(\omega'_1, \omega'_2) &=& a(\omega'_1) +
\frac{b(\omega'_1)}{\omega'_2} + \dots
\end{eqnarray}
where the further terms in the expansion of 
$f(\omega'_1, \omega'_2)$ for $\omega'_2 \to \infty$
lead to convergent expressions when
integrated over $\omega'_2$ in the region of large $\omega'_2$.
The leading coefficient is
\begin{equation}
a(\omega'_1) = \omega'_1\,\left< p'^i \, \frac{H'-E'}{(H'-E'+\omega'_1)^2}
\, p'^i \right> \,,
\end{equation}
and the second reads
\begin{equation}
\label{bofomega1}
b(\omega'_1) = \omega'_1 \,\, \delta_{W} \left\{ \left<
p'^i \, \frac{1}{E'-(H'+\omega'_1)}\, p'^i \right>  \right\} \,,
\end{equation}
where by $\delta_{W}$ we denote the first-order correction
to the quantity in curly brackets obtained via the action of the 
scaled, dimensionless, local potential 
\begin{equation} 
\label{deltaW}
W = \frac{\pi \, \delta^{(3)}(\bm{r})}{(Z\alpha)^3\,m^3}
= \pi \, \delta^{(3)}(\bm{r}')\,,
\end{equation}
i.e., by the replacements [see Eq.~(\ref{defG})],
\begin{subequations}
\begin{eqnarray}
H' &\to& H' + W\,, \\[2ex]
|\phi\rangle &\to& |\phi\rangle + |\delta \phi\rangle\,, \\[2ex]
E' &\to& E' + \delta E'\,.
\end{eqnarray}
\end{subequations}
Here
\begin{equation}
\label{deltaE}
\delta E' = \langle W \rangle \,, \qquad
| \delta \phi \rangle = G'_{\rm red}(0) \, W | \phi \rangle\,.
\end{equation}
The correction (\ref{bofomega1})
has been calculated for excited $S$ states in
Ref.~\cite{Je2003jpa}.

%
%
\begin{figure}[htb]%
\begin{center}\includegraphics[width=0.9\linewidth]{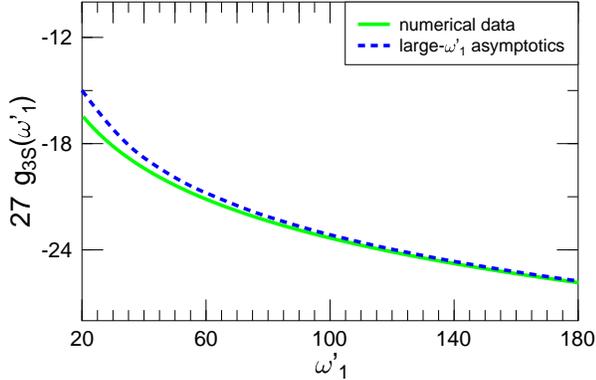}\end{center}%
\caption{\label{fig3}%
(color online).          
Plot of the large-$\omega'_1$ asymptotics of $g$ 
[see Eq.~(\ref{asymptoticsg})] against 
numerical data obtained for $g_{3S}(\omega'_1)$ in the 
range $\omega'_1 \in [20, 180]$. 
The numerical data are scaled by a factor of $3^3 = 27$.
See also Table~\ref{table1} and Eq.~(\ref{defg}) 
where $g$ is defined. Dimensionless quantities
are displayed in the figure [this statement 
relates to both the abscissa as well
as the ordinate axis, see also Eq.~(\ref{scaling})].}
\end{figure}

We are interested in evaluating the constant term 
$g(\omega'_1)$ in the 
integral of $f(\omega'_1, \omega'_2)$ in the 
range $\omega'_2 \in (0, \Lambda)$ at fixed $\omega'_1$ for large
$\Lambda$:
\begin{subequations}
\label{funcg}
\begin{equation}
\label{defg}
\int_0^\Lambda {\rm d}\omega'_2 \,
f(\omega'_1, \omega'_2) = a(\omega'_1) \, \Lambda + 
b(\omega'_1) \, \ln \Lambda + g(\omega'_1)\,,
\end{equation}
where we neglect terms that vanish as $\Lambda \to \infty$.
This equation provides an implicit definition of
$g(\omega'_1)$ as the constant term which results
in the limit $\Lambda \to 0$. The constant term may be evaluated as
\begin{equation}
\label{evalg}
g(\omega'_1) = {\cal I}_1 + {\cal I}_2 + {\cal I}_3\,,
\end{equation}
\end{subequations}
where
\begin{subequations}
\label{I}
\begin{eqnarray}
\label{I1}
{\cal I}_1 &=& \int_0^M {\rm d}\omega'_2 \, f(\omega'_1, \omega'_2) \,,
\\[2ex]
\label{I2}
{\cal I}_2 &=& \int_M^\infty {\rm d}\omega'_2  
\left[ f(\omega'_1, \omega'_2) - a(\omega'_1) -
\frac{b(\omega'_1)}{\omega'_2} \right]\,, \\[2ex]
\label{I3}
{\cal I}_3 &=& -a(\omega'_1) \, M -b(\omega'_1) \, \ln M\,,
\end{eqnarray}
\end{subequations}
with arbitrary $M$ [the result for $g(\omega'_1)$ is
independent of $M$].
Sample values of the $g$-function for $nS$ states
are given in Table~\ref{table1}. The sign of the ${\cal I}_3$-term 
(cf.~\cite[Eq.~(8c)]{PaJe2003})
is determined by the necessity of subtracting the integral 
of the subtraction term [second term in the integrand 
of Eq.~(\ref{I2})], 
at the lower limit of integration $M$.
In both the $\omega'_1$ as well as the $\omega'_2$
integrations, there is a further complication due to bound-state poles
($P$ states) which need to be considered for higher 
excited $nS$ states ($n \geq 3$).
In the current section, we completely ignore the imaginary parts and 
carry out all integrations with a principal-value prescription.
{\em Idem est}, we use the prescription ($M > a$)
\begin{equation}
\label{prec1}
\int_0^M {\rm d} \omega' \, \frac{1}{(\omega' - a)} \to
\ln\left(\frac{M - a}{a}\right)\,.
\end{equation}
For double poles, which originate from some of the terms
in Eq.~(\ref{NRQED}), the
appropriate integration prescription is as follows:
\begin{equation}
\label{prec2}
\int_0^M {\rm d} \omega' \, \frac{1}{(\omega' - a)^2} \to 
\frac{M}{a\, (a-M)}\,.
\end{equation}
Even if $M > a$, this prescription leads to a finite result
which is real rather than complex. 
The same result can also be obtained under a
symmetric deformation of the integration 
contour into the complex plane. Analogous integration prescriptions
have been used in~\cite{JePa1996,JeSoMo1997}.
Double poles normally lead to 
nonintegrable singularities and give rise to 
serious concern. It is therefore necessary 
to ask how these terms originate in the context of the 
current calculation.
To answer this question it is useful to remember
that expression~(\ref{NRQED}) is obtained
by perturbation theory in powers of the nonrelativistic QED interaction
Lagrangian; an expansion in powers of the
interaction is, however, not allowed
when we are working close
to a resonance of the unperturbed atomic Green 
function---i.e.,~close to a bound-state pole. 
The double poles incurred by this expansion 
find a natural {\em a posteriori} treatment
by the prescriptions (\ref{prec1}) and (\ref{prec2}) above.
In general, double poles as encountered here and
previously in~\cite{JePa1996,JeSoMo1997} originate
whenever we work with {\em (i)} excited states which can decay through
$E1$ one-photon emission and {\em (ii)} propagators are 
perturbatively expanded near bound-state poles.

%
%
\begin{figure}[htb]%
\begin{center}
\begin{center}\includegraphics[width=0.9\linewidth]{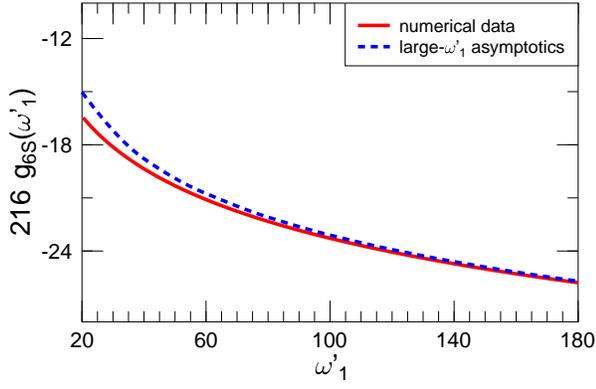}\end{center}%
\end{center}%
\caption{\label{fig4}%
(color online). 
Large-$\omega'_1$ asymptotics of $g$ plotted against numerical data for 
the $6S$ state in the range $\omega'_1 \in [20, 180]$. 
The numerical data are scaled by a factor of $6^3 = 216$.
Explicit numerical sample values 
for $g_{6S}(\omega'_1)$ can also be found in
Table~\ref{table1}. The apparent similarity of Figs.~\ref{fig3}
and~\ref{fig4} is reflected in the scaled entries in this Table.
The difference between the numerical data (solid line) and 
the asymptotics (dashed line) is negative.
The formula for the large-$\omega'_1$ asymptotics of
$g$ is given in Eqs.~(\ref{asymptoticsg}) and~(\ref{asymptoticsgcoeff}).
The difference between the numerical data and 
the asymptotics gives rise to a negative contribution 
to the integral $J_2$ defined according to Eq.~(\ref{J2})
and to a negative value for the $B_{60}$-coefficient
[see Eq.~(\ref{b60s}) below]. 
The scaled, primed quantities plotted along both the abscissa as well as the 
ordinate axis are dimensionless
[see also Eq.~(\ref{scaling})].}%
\end{figure}

The leading terms in the asymptotics of $g(\omega'_1)$ for 
large $\omega'_1$ read~\cite{PaJe2003}
\begin{eqnarray}
\label{asymptoticsg}
g(\omega'_1) &=& 
\frac{1}{n^3} \, \left[ A\, \ln\omega'_1 + B + 
C \, \frac{\ln(\omega'_1)}{\sqrt{\omega'_1}} +
D \, \frac{1}{\sqrt{\omega'_1}} \right.
\nonumber\\[2ex]
& & \left. + E \, \frac{\ln^2(\omega'_1)}{\omega'_1} +
F \, \frac{1}{\omega'_1} \right] + \dots,
\end{eqnarray}
where
\begin{subequations}
\label{asymptoticsgcoeff}
\begin{eqnarray}
A &=& - 4\,,\\[2ex]
B &=& 2\,\left[\ln 2 - 1 - \ln k_0(nS)\right] \,,\\[2ex]
C &=& 4 \, \sqrt{2}\,,\\[2ex]
D &=& 4 \, \sqrt{2}\, \left( 2 \, (\ln 2 - 1) - \pi\right) \,,\\[2ex]
E &=& 1\,,\\[2ex]
F &=& 8 + \frac{3}{2} \, N(nS) + 5 \pi^2\,.
\end{eqnarray}
\end{subequations}
The higher-order terms in the large-$\omega'_1$ expansion, which are ignored in
Eq.~(\ref{asymptoticsg}), lead to convergent expressions in the 
problematic integration region $\omega'_1 \to \infty$.
Explicit numerical values for $N(nS)$ are given in Eq.~(\ref{resultsnS}).
For ${3S}$ and $6S$ states, numerical data for $g$ are compared to the 
leading asymptotics in Figs.~\ref{fig3} and~\ref{fig4}.

\begin{figure}[htb]%
\begin{center}
\begin{center}\includegraphics[width=0.9\linewidth]{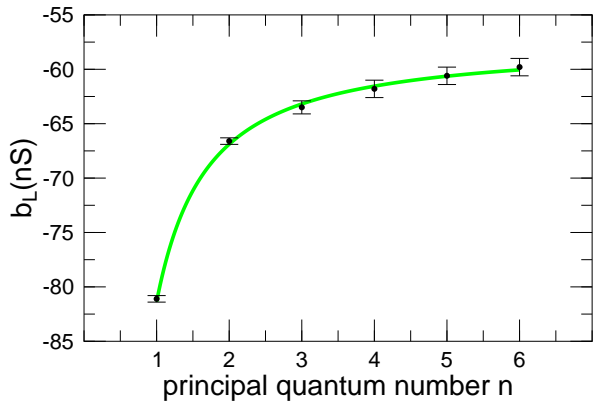}\end{center}%
\end{center}%
\caption{\label{fig5}%
(color online). Dependence of the two-loop Bethe logarithm $b_L(nS)$
on the principal quantum number $n$. The explicit
numerical results [Eq.~(\ref{bL})] are displayed together
with a three-parameter fit of the form 
$-57.4 - 13.7/n - 10.1/n^2$
from which one may infer 
$\lim_{n\to\infty} b_L(nS) = -57(2)$. 
Quantities plotted along the abscissa and the ordinate axis are
(of course) dimensionless.}%
\end{figure}

The two-loop Bethe logarithm, which
is equal to the low-energy contribution 
$B^{\rm lep}_{60}(nS)$ to the coefficient
$B^{\rm (2LSE)}_{60}$ [see Eq.~(\ref{defH})], 
can be obtained by considering
\begin{equation}
\int_0^\Lambda {\rm d}\omega'_1 \, g(\omega'_1)\,,
\end{equation} 
and subtracting all terms that diverge as $\Lambda \to \infty$,
as given by the leading asymptotics in Eq.~(\ref{asymptoticsg}).
Specifically, the integration procedure is as follows. We define
the two-loop Bethe logarithm as
\begin{equation}
b_L(nS) = n^3 \, \left( {\cal J}_1 + {\cal J}_2 + {\cal J}_3\right)\,,
\end{equation}
where
\begin{subequations}
\label{J}
\begin{eqnarray}
\label{J1}
{\cal J}_1 &=& \int_0^N {\rm d}\omega'_1 \, g(\omega'_1) \,,\\[2ex]
\label{J2}
{\cal J}_2 &=& \int_N^\infty {\rm d}\omega'_1 \, 
\left[ g(\omega'_1) - \frac{1}{n^3} \,
\biggl( A\, \ln\omega'_1 + B 
\right. 
\nonumber\\[2ex]
& & + C \, \frac{\ln(\omega'_1)}{\sqrt{\omega'_1}} 
+ D \, \frac{1}{\sqrt{\omega'_1}} 
\nonumber\\[2ex]
& & \left. + E \, \frac{\ln^2(\omega'_1)}{\omega'_1} +
F \, \frac{1}{\omega'_1} \biggr) \right]\,,
\\[2ex]
\label{J3}
{\cal J}_3 &=& A \, N (\ln N - 1) + B\,N 
\nonumber\\[2ex]
& & + C\, \sqrt{N} \, \left(\ln N - 2\right) 
+ 2 D \, \sqrt{N}
\nonumber\\[2ex]
& & + 2 E \, \sqrt{N} \, \left(8 + (\ln N - 4) \, \ln N \right)
\nonumber\\[2ex]
& & + F\, \ln N\,.
\end{eqnarray}
\end{subequations}
Again, in analogy to the integration prescription
in Eqs.~(\ref{funcg}) and (\ref{I}),
the result for $b_L$ is independent of the choice of $N$.
Our numerical results for the two-loop Bethe logarithm of 1S and 2S
states read (results for $1S$ and $2S$ are quoted from~\cite{PaJe2003}):
\begin{subequations}
\label{bL}
\begin{eqnarray}
b_L(1S) &=& 
-81.4(3)\,, \\[2ex]
b_L(2S) &=& 
-66.6(3)\,, \\[2ex]
b_L(3S) &=& 
-63.5(6)\,, \\[2ex]
b_L(4S) &=& 
-61.8(8)\,, \\[2ex]
b_L(5S) &=& 
-60.6(8)\,, \\[2ex]
b_L(6S) &=& 
-59.8(8)\,.
\end{eqnarray}
\end{subequations}
These results are displayed in Fig.~\ref{fig5}.

From here on,
we restore in the following formulas the physical dimensions of all energies
and frequencies and revoke the scaling introduced in
Eq.~(\ref{scaling}). Primed quantities will no longer be
used in the following sections of this work.

%
%

%
%
\section{AMBIGUITY IN THE DEFINITION OF $B_{60}$}
\label{ambiguity}

Low~\cite{Lo1952} was the first to point out that 
the definition of an atomic energy level becomes problematic
at the order of $\alpha^8$ [more specifically, $\alpha^2\,(Z\alpha)^6\,m$],
and that it becomes necessary at this level of accuracy to consider
the contribution of nonresonant energy levels to the elastic 
scattering cross section. In~\cite{JeMo2002}, it has been stressed that
nonresonant effects are enhanced in differential as opposed 
to total cross section, leading to corrections of order 
$\alpha^2\,(Z\alpha)^4\,m$. Related issues have recently attracted
some attention (see also the discussion in Sec.~\ref{intro}), and 
there is even a connection to the two-loop corrections
of order $\alpha^2\,(Z\alpha)^6\,m$, as we discuss in the following. 
Namely, as pointed out in~\cite{JeEvKePa2002}, 
the two-loop self-energy contains contributions which result
from squared decay rates. 

For excited reference states,
the nonrelativistic two-loop self-energy (\ref{NRQED}) contains
imaginary contributions which are generated by both the $\omega_1$- as well
as the $\omega_2$-integrations. The imaginary part of the 
one-photon self-energy
is generated by a pole contribution and leads to the decay rate 
which is the imaginary part of the self-energy.
Consequently, real contributions 
to the two-photon self-energy which result as a product of two
imaginary contributions are naturally referred to as squared decay rates.
These are natural contributions to the two-loop self-energy shift
in the order of $\alpha^2\,(Z\alpha)^6\,m$ and cannot be associated
in a unique manner with one and only one atomic level.
Roughly speaking, the problems
in the interpretation originate from the fact that the
Gell--Mann--Low--Sucher~\cite{GMLo1951,Su1957} 
formalism involves {\em a priori}
asymptotic states with an infinite lifetime (vanishing decay
rate). Furthermore, it has been mentioned~\cite{JeKePa2002} that 
problematic issues persist even if the concept of an
atomic energy is generalized to a resonance with a finite 
width---i.e.,~even if the canonical concept of a pole of the 
resolvent on the second Riemann sheet~\cite{CTDRGr1989,CTDRGr1992} 
is used for the definition of an atomic resonance.
In general, the squared decay rates
illustrate that we are reaching the limit of the proper definition 
of atomic energy levels in considering higher-order
two-loop binding corrections.

In~\cite{JeEvKePa2002}, the squared decay rates have been analyzed
in some detail. There are four specific terms 
out of the nine in curly brackets in Eq.~(\ref{NRQED})
which give rise to squared decay rates. We list these terms
here with a special emphasis on the higher excited $3S$ state,
following the notation introduced in~\cite{JeEvKePa2002}.
In the following formulas,
the physical dimensions of all energies
and frequencies are restored [cf.~Eq.~(\ref{scaling})], 
and we have for the first term ${\mathcal T}_1({3S})$,
which is the analog of Eq.~(4)~of~\cite{JeEvKePa2002}:
\begin{widetext}
\begin{subequations}
\begin{eqnarray}
\lefteqn{
\label{T1}
{\mathcal T}_1({3S})
= \lim_{\delta \to 0^+} - \left( \frac{2 \, \alpha}{3 \,\pi\,m^2} \right)^2 \,
\int_0^{\epsilon_1} {\rm d}\omega_1 \, \omega_1 \,
\int_0^{\epsilon_2} {\rm d}\omega_2 \, \omega_2 \,} \nonumber\\[2ex]
& & 
\times 
\left< {3S} \left|  p^i \, 
\frac{1}{H - {\rm i}\,\delta - E_{3S} + \omega_1} \, p^j \,
\frac{1}{H - E_{3S} + \omega_1 + \omega_2} \, p^i \,
\frac{1}{H - {\rm i}\,\delta - E_{3S} + \omega_2} \, 
p^j \right| {3S} \right>\,.
\end{eqnarray}
For the analog of Eq.~(8)~of~\cite{JeEvKePa2002}, we have
\begin{eqnarray}
\label{T2}
\lefteqn{{\mathcal T}_2({3S})
= \lim_{\delta \to 0^+}
- \left( \frac{2 \, \alpha}{3 \,\pi\,m^2} \right)^2 \,
\int_0^{\epsilon_1} {\rm d}\omega_1 \, \omega_1 \,
\int_0^{\epsilon_2} {\rm d}\omega_2 \, \omega_2 \,} \nonumber\\[2ex]
& & \times
\left< {3S} \left|  p^i \, 
\frac{1}{H - {\rm i}\, \delta - E_{3S} + \omega_1} \, p^i \,
\left( \frac{1}{H - E_{3S}} \right)' \, p^j \,
\frac{1}{H - {\rm i}\, \delta - E_{3S} + \omega_2} \, p^j 
\right| {3S} \right>\,, 
\end{eqnarray}
and we also have [see Eq.~(15) of~\cite{JeEvKePa2002}]
\begin{eqnarray}
\label{T3}
\lefteqn{{\mathcal T}_3({3S}) 
= \lim_{\delta \to 0^+}
\left( \frac{2 \, \alpha}{3 \,\pi\,m^2} \right)^2 \,
\int_0^{\epsilon_1} {\rm d}\omega_1 \, \omega_1 \,
\int_0^{\epsilon_2} {\rm d}\omega_2 \, \omega_2 \,} \nonumber\\[2ex]
& &
\times
\left< {3S} \left|  p^i \, 
\frac{1}{H - {\rm i}\,\delta - E_{3S} + \omega_1} \, p^i \,
\right| {3S} \right> \,
\left< {3S} \left|  p^j \, 
\left( \frac{1}{H - {\rm i}\,\delta - E_{3S} + \omega_2} \right)^2 \, 
p^j \right| {3S} \right>\,.
\end{eqnarray}
The last relevant term is 
[see Eq.~(17) of~\cite{JeEvKePa2002}]
\begin{eqnarray}
\label{T4}
\lefteqn{{\mathcal T}_4({3S})
= \lim_{\delta \to 0^+} \left( \frac{2 \, \alpha}{3 \,\pi\,m^2} \right)^2 \,
\int_0^{\epsilon_1} {\rm d}\omega_1 \, \omega_1 \,
\int_0^{\epsilon_2} {\rm d}\omega_2 \, \omega_2 \,} \nonumber\\[2ex]
& &
\times
\left< {3S} \left|  p^i \, 
\frac{1}{H - {\rm i}\,\delta - E_{3S} + \omega_1} \,
\frac{1}{H - {\rm i}\,\delta - E_{3S} + \omega_2} \, 
p^i \right| {3S} \right>\,.
\end{eqnarray}
\end{subequations}
\end{widetext}
Here, $H$ is the Schr\"{o}dinger Hamiltonian. 
We now proceed to analyze the squared decay rates
generated by the terms ${\cal T}_i$ ($i=1,\dots,4$) in some detail.
It should be reemphasized here that the main contributions
to the energy shift generated by the ${\cal T}_i$ have already been
analyzed in Sec.~\ref{2Lnrqed}. However, the prescriptions
(\ref{prec1}) and (\ref{prec2}) lead to a complete neglect
of the (squared) imaginary contributions. Consequently,
we here ``pick up'' only the terms of the 
``squared-decay'' type---i.e.,~the terms
generated by the infinitesimal half-circles around the poles
at $\omega_1 = E_{3S} - E_{2P}$ and 
$\omega_2 = E_{3S} - E_{2P}$. For the evaluation
of these squared pole terms, specification of the infinitesimal imaginary
part $- {\rm i}\,\delta$ is required in order to fix the sign 
of the pole contribution. This procedure of extracting 
squared imaginary parts leads to the 
terms ${\cal C}_i$ ($i=1,\dots,4$), respectively~\cite{remark3S}.

We now proceed to analyze the squared decay rates 
generated by the terms ${\cal T}_i$ ($i=1,\dots,4$) in some detail.
The term ${\cal T}_1$ is due to the diagram with crossed loops
in Fig.~\ref{fig1}(A). For the contribution $C_1({3S})$ 
generated by the poles at  $\omega_1 = E_{3S} - E_{2P}$ and 
$\omega_2 = E_{3S} - E_{2P}$ in ${\mathcal T}_1({2P})$, 
we obtain
\begin{eqnarray}
\label{C1}
C_1({3S}) & = & \alpha^2 \, \frac{4}{9 m^4} 
\left(E_{3S} - E_{2P}\right)^2 \, 
  |\langle {2P} | \bm{p} | {3S} \rangle|^2 
\nonumber\\[2ex]
& & \times \left< {2P} \left| p^i \,   
       \frac{1}{H + E_{3S} - 2\,E_{2P}} \, p^i 
          \right| {2P} \right>
\nonumber\\[3ex]
& = & \frac{2^5}{3^3 \,5^8} \, \alpha^2 \, (Z\alpha)^6 \, m \, {\cal M}_1\,,
\end{eqnarray}
where we define
$| {2P} \rangle$ to be the state with magnetic quantum number
(angular momentum projection) $m=0$.
This explains the 
additional factor of $3$ in comparison
to Eq.~(5) of~\cite{JeEvKePa2002}.
The factor originates from the summation over magnetic quantum 
numbers of the $| {2P} \rangle$ state, and we reemphasize 
that we understand by
$| {2P} \rangle$ only the state with magnetic quantum number 
(angular momentum projection) $m=0$.
The matrix element ${\cal M}_1$ reads
\begin{equation}
{\cal M}_1 = \frac{1}{m} \,     
  \left< {2P} \left| \, p^i \,
    \frac{1}{H + E_{3S} - 2\,E_{2P}} \, p^i \,
      \right| {2P} \right> = 0.697\,,
\end{equation}
and we have for the well-known dipole matrix element
\begin{equation}
\left| \left< {2P} 
\left| \frac{\bm{p}}{m} \right| 
{3S} \right> \right|^2 = 
\frac{2^{9}\,3^3}{5^{10}} \, (Z\alpha)^2\,.
\end{equation}
Note that the contribution $C_1$ lacks the factors $\pi$ in the denominator
which are characteristic of other two--loop corrections: these are compensated
by additional factors of $\pi$ in the numerator that characterize the 
pole contributions.

The rainbow diagram in Fig.~\ref{fig1}(B) 
with the second loop inside the first 
does not create squared imaginary contributions.
From the irreducible part of the
loop-after-loop diagram in Fig.~\ref{fig1}(C) (we exclude the reference state
in the intermediate electron propagator), the term ${\cal T}_2$
is obtained.
Again, picking up only those terms which are
generated by the infinitesimal half-circles around the poles
at $\omega_1 = E_{3S} - E_{2P}$ and
$\omega_2 = E_{3S} - E_{2P}$,
we obtain the contribution $C_2({3S})$ involving
squared decay rates:
\begin{eqnarray}
\label{C2}
C_2({3S}) & = & \alpha^2 \, \frac{4}{9 m^4}
\left(E_{3S} - E_{2P}\right)^2 \,
  |\langle {2P} | \bm{p} | {3S} \rangle|^2 
\nonumber\\[2ex]
& & \times \left< {2P} \left| p^i \,
       \left( \frac{1}{H - E_{3S}} \right)' \, p^i
          \right| {2P} \right>
\nonumber\\[3ex]
& = & \frac{2^5}{3^3\,5^8}\, \alpha^2 \, (Z\alpha)^6 \, m \, {\cal M}_2\,,
\end{eqnarray}
where the matrix element ${\cal M}_2$ reads
\begin{equation}
{\cal M}_2 = \frac{1}{m} \,
  \left< {2P} \left| \, p^i \,
    \left( \frac{1}{H - E_{3S}} \right)' \, p^i \,
      \right| {2P} \right> = 0.490\,.
\end{equation}
The prime in the reduced Green function
indicates that the $3S$ state is excluded from the
sum over intermediate states, and it should not 
be confused with the notation used 
in Sec.~\ref{2Lnrqed}, where the prime was used to 
denote scaled dimensionless instead of 
dimensionful quantities. 

From the derivative term (reducible part of the loop-after-loop
diagram), we obtain
\begin{eqnarray}
\label{C3}
C_3({2P}) & = & - \alpha^2 \, \frac{4}{3 m^4}
\left(E_{3S} - E_{2P}\right) \,
  |\langle {2P} | \bm{p} | {3S} \rangle|^4 \,
\nonumber\\[2ex]
&=& - \frac{2^{17}\,3^3}{5^{19}} \, \alpha^2 \, (Z\alpha)^6 \, m \,.
\end{eqnarray}
In order to derive the imaginary parts, one should remember that 
the squared propagator originates from a differentiation 
of a single propagator with respect to the energy. 
An integration by parts is helpful.

The last contribution of the ``squared-decay'' 
type---it originates from the ``seagull term'' characteristic of
NRQED---is ${\cal T}_4$. The corresponding ${\cal C}$-term is
\begin{eqnarray}
\label{C4}
C_4({3S}) & = & - \alpha^2 \, \frac{4}{3 m^3}
\left(E_{3S} - E_{2P}\right)^2 \,
  |\langle {2P} | \bm{p} | {3S} \rangle|^2 \,
\nonumber\\[3ex]
& = & - \frac{2^{5}}{3^2\,5^{8}} \, \alpha^2 \, (Z\alpha)^6 \, m\,.
\end{eqnarray}
Adding all contributions, we obtain a shift of
\begin{equation}
\label{C3S}
\sum_{i=1}^4 C_i({3S}) = \left(\frac{\alpha}{\pi}\right)^2 \,
\frac{(Z\alpha)^6 \, m}{3^3} \, (-0.00151)
\end{equation}
for the $3S$ level. The numerical value is tiny; 
for $Z=1$ the shift amounts to only
\begin{equation}
\label{delta2nu3s}
\delta^2\nu(3S) = -0.00565\,{\rm Hz}\,. 
\end{equation}
For the corresponding ambiguous contributions
to the $B_{60}$-coefficient [see Eqs.~(\ref{defH}) and~(\ref{C3S})],
we use the notation 
\begin{equation}
\label{delta23S}
\delta^2 B_{60}(3S) = -0.00151\,. 
\end{equation}
For the $4S$ state, we have to take into account the decays
into the $2P$ and $3P$ states. For example, the 
contribution $C_1(4S)$ reads
\begin{widetext}
\begin{eqnarray}
C_1(4S) & = & 
\alpha^2 \, \frac{4}{9 m^4} \, \left\{
\left(E_{\rm 4S} - E_{2P}\right)^2 \, 
  |\langle {2P} | \bm{p} | {\rm 4S} \rangle|^2 \,
    \left< {2P} \left| p^i \,   
       \frac{1}{H + E_{\rm 4S} - 2\,E_{2P}} \, p^i 
          \right| {2P} \right> \right. 
\nonumber\\[1ex]
& & \quad \left. 
+ \left(E_{\rm 4S} - E_{\rm 3P}\right)^2 \, 
  |\langle {\rm 3P} | \bm{p} | {\rm 4S} \rangle|^2 \,
    \left< {\rm 3P} \left| p^i \,   
       \frac{1}{H + E_{\rm 4S} - 2\,E_{\rm 3P}} \, p^i 
          \right| {\rm 3P} \right> \right\}
\nonumber\\[1ex]
& & + \alpha^2 \, \frac{8}{9 m^4} \, 
(E_{\rm 4S} - E_{2P}) \, (E_{\rm 4S} - E_{\rm 3P}) \,
{\rm Re}\left(\langle {\rm 4S} | p^j | {2P} \rangle \,
\left< {2P} \left| p^i \,   
  \frac{1}{H + E_{\rm 4S} - E_{\rm 3P} - E_{2P}} \, p^i 
     \right| {\rm 3P} \right> \,
\langle {\rm 3P} | p^j | {\rm 4S} \rangle
\right)
\nonumber\\[3ex]
& = & \left(\frac{\alpha}{\pi}\right)^2 \,
\frac{(Z\alpha)^6 \, m}{4^3} \, (0.00108)\,.
\end{eqnarray}

\begin{center}
\begin{table}
\begin{center}
\begin{minipage}{11.7cm}
\caption{\label{table1}
Sample values of the $g$ function, defined in Eq.~(\ref{funcg}),
for the $nS$ states with $n=1,\dots,6$. Multiplication by a 
factor of $n^3$ approximately accounts for the $n$-dependence,
in agreement with the $n^{-3}$-type scaling of the two-loop
correction as implied by Eq.~(\ref{DefESE2L}).} 
\begin{tabular}{r@{\hspace{0.3cm}}r@{\hspace{0.3cm}}r@{\hspace{0.3cm}}r%
@{\hspace{0.3cm}}r@{\hspace{0.3cm}}r@{\hspace{0.3cm}}r}
\hline
\hline
\rule[-3mm]{0mm}{8mm}
$\omega$ & $g_{1S}$ & $8\,g_{2S}$ & $27\,g_{{3S}}$ & $64\,g_{4S}$%
& $125\,g_{{3S}}$ & $216\,g_{6S}$ \\
\hline
\rule[-3mm]{0mm}{8mm}
   0  &   0.000\,00 &   0.000\,00  &    0.000\,0 &   0.000\,0%
&  0.000\,0 &  0.000\,0\\
\rule[-3mm]{0mm}{8mm}
   5  & -10.281\,60 & -10.367\,94  &  -10.450\,1 & -10.490\,8%
& -10.522\,6 & -10.546\,0\\
\rule[-3mm]{0mm}{8mm}
  20  & -16.560\,34 & -16.415\,97  &  -16.393\,4 & -16.385\,1%
& -16.386\,1 & -16.386\,7\\
\rule[-3mm]{0mm}{8mm}
  80  & -22.714\,02 & -22.439\,66  &  -22.372\,0 & -22.345\,5%
& -22.332\,0 & -22.326\,3\\
\rule[-3mm]{0mm}{8mm}
 180  & -26.232\,35 & -25.923\,09  &  -25.848\,0 & -25.813\,6%
& -25.798\,1 & -25.789\,5 \\
\hline
\hline
\end{tabular}
\end{minipage}
\end{center}
\end{table}
\end{center}

\begin{center}
\begin{table}
\begin{center}
\begin{minipage}{17.3cm}
\caption{\label{table2} Squared decay rates are extracted 
as the squared bound-state pole terms from the
terms ${\cal T}_1$---${\cal T}_4$ in Eqs.~(\ref{T1})---(\ref{T4}).
Explicit formulas ($3S$ state) for 
the terms ${\cal C}_i$ ($i=1,\dots,4$) are given in
Eqs.~(\ref{C1})---(\ref{C4}).
All contributions ${\cal C}_i$ scale as $Z^6$, 
whereas the decay rates $\Gamma$ given in the eighth
column scale as $Z^4$. Numerical values are 
given for $Z=1$. The decay rates 
may be derived in the standard way
[see the derivation of Fermi's golden rule as 
given in Eqs.~(2.103)---(2.118) of~\cite{Sa1967Adv}]. 
We only indicate approximate values for $\Gamma$,
without relativistic corrections.
For the $2P_{1/2}$ states, a detailed calculation
leads to $\Gamma(2P_{1/2}) = 0.9970942\,Z^4\,{\rm MHz}$~\cite{SaPaCh2004}.
For any given state, the squared decay rates $\delta^2 \nu$ 
are about seven to eight orders of magnitude
smaller than the width $\Gamma$.
All states listed in the 
table may decay via the $E1$ mode,
wherefore the decay rates as well as the ambiguities
$\delta^2\nu$ are formally of the same 
order-of-magnitude [$\alpha\,(Z\alpha)^4\,m$ and
$\alpha^2\,(Z\alpha)^6\,m$, respectively].
However, the numerical coefficients differ by two orders of magnitude;
$S$ states typically have a much longer lifetime.}
\begin{tabular}{c@{\hspace{0.3cm}}r@{\hspace{0.3cm}}r@{\hspace{0.3cm}}%
r@{\hspace{0.3cm}}r@{\hspace{0.3cm}}%
r@{\hspace{0.3cm}}r@{\hspace{0.3cm}}r@{\hspace{0.3cm}}r}
\hline
\hline
\rule[-3mm]{0mm}{8mm} state &
\multicolumn{1}{c}{${\cal C}_1$} & 
\multicolumn{1}{c}{${\cal C}_2$} & 
\multicolumn{1}{c}{${\cal C}_3$} & 
\multicolumn{1}{c}{${\cal C}_4$} &
\multicolumn{1}{c}{$\delta^2\nu = \sum_{i=1}^4 {\cal C}_i$} &
\multicolumn{1}{c}{$\delta^2 B_{60}$} &
\multicolumn{1}{c}{$\Gamma$} &
\multicolumn{1}{c}{$\tau$} \\
\hline
\rule[-3mm]{0mm}{8mm} ${2P}$ &
$1.42208$ Hz &
$2.06790$ Hz &
$-1.00843$ Hz &
$-4.84593$ Hz &
$-2.36438$ Hz &
$-0.18789$ &
$99.76$ MHz &
$0.16\times10^{-8}$ s\\
\rule[-3mm]{0mm}{8mm} $3P$ &
$0.50353$ Hz &
$0.06037$ Hz &
$-0.12787$ Hz &
$-2.00952$ Hz &
$-1.57349$ Hz &
$-0.42202$ &
$30.21$ MHz &
$0.53\times10^{-8}$ s \\
\rule[-3mm]{0mm}{8mm} ${3S}$ &
$0.00210$ Hz &
$0.00148$ Hz &
$-0.00018$ Hz &
$-0.00565$ Hz &
$-0.00564$ Hz &
$-0.00151$ &
$1.00$ MHz &
$15.83\times10^{-8}$ s \\
\rule[-3mm]{0mm}{8mm} $4S$ &
$0.00170$ Hz &
$-0.00100$ Hz &
$-0.00015$ Hz &
$-0.01019$ Hz &
$-0.00964$ Hz &
$-0.00613$ &
$0.70$ MHz &
$22.65\times10^{-8}$ s\\
\hline
\hline
\end{tabular}
\end{minipage}
\end{center}
\end{table} 
\end{center} 
\end{widetext}

The sum of $C_{1,\dots,4}$ for the $4S$ state of hydrogenlike 
systems with (low) nuclear charge $Z$ is
\begin{equation}
\label{res4Sstate}
\sum_{i=1}^4 C_i(4S) = \left(\frac{\alpha}{\pi}\right)^2 \,
\frac{(Z\alpha)^6 \, m}{4^3} \, (-0.00613)\,.
\end{equation}
For atomic hydrogen ($Z=1$), this correction evaluates to 
\begin{equation}
\label{delta2nu4s}
\delta^2\nu(4S) = -0.00964\,{\rm Hz}\,. 
\end{equation}
This is numerically larger than the corresponding effect for ${3S}$
[see Eq.~(\ref{delta2nu3s})].
We have 
\begin{equation}
\label{delta24S}
\delta^2 B_{60}(4S) = -0.00613\,.
\end{equation}
Although self-energy corrections canonically
scale as $n^{-3}$ [see Eq.~(\ref{DefESE2L})], 
the coefficient in this case grows so rapidly 
with $n$ that the correction is enhanced for $4S$ in comparison
to $3S$. Further detailed information can be found in 
Table~\ref{table2}. We also take the opportunity to clarify that
numerical values for squared decay as given in~\cite{JeEvKePa2002}
(for $2P$ and $3P$ states)
should be understood as given in inverse seconds rather than Hz
(see also the discussion near the beginning of Sec.~\ref{2Lnrqed}).

%
%
\section{FURTHER CONTRIBUTIONS TO $B_{60}$}
\label{further}

The coefficient $B_{60}$ can be represented as the sum
\begin{equation}
B_{60} = b_L+b_{M}+b_{F}+b_{H}+b_{\rm VP}\,.
\end{equation}
The two-loop Bethe logarithm $b_L$
comes from the region where both photon momenta are small
and has been the subject of this work.
$b_{M}$ stems from an integration region
where one momentum is large $\sim m$, and the second
momentum is small.
This contribution is given by a Dirac $\delta$ correction to
the Bethe logarithm [see also Eq.~(\ref{bofomega1}) 
and Ref.~\cite{Je2003jpa}]. It
has already been derived in \cite{Pa2001} but not included
in the theoretical predictions for the Lamb shift:
\begin{equation}
\label{bM}
b_{M} = \frac{10}{9}\,N(nS)\,.
\end{equation}
As has already been mentioned in~\cite{PaJe2003},
the contributions
$b_{F}$ and $b_{H}$ originate from a region where both photon
momenta are large $\sim m$, and the electron momentum is small and
large respectively. Finally, $b_{\rm VP}$ is a contribution from
diagrams that involve a closed fermion loop. None of these effects
have been calculated as yet. On the basis of our experience with
the one- and two-loop calculations 
we estimate the magnitude of
these uncalculated terms to be of the order of 15\%. 
For higher excited states
($3S,\dots,6S$), the $15\,\%$ uncertainty due to unknown
contributions is larger than the ambiguity $\delta^2 B_{60}$
listed in Table~\ref{table2}.

Concerning logarithmic two-loop vacuum-polarization effects
\cite{Pa2001}, we mention that 
the contribution of the two-loop self-energy diagrams
to $B_{61}$ for the $1S$ state reads $49.8$, whereas the diagrams
that involve a closed fermion loop amount to $0.6$. Concerning
the one-loop higher-order binding correction $A_{60}(1S)$ (analog
of $B_{60}$) it is helpful to consider that 
the result for $1S$ is $-30.92415(1)$
(see Refs.~\cite{Pa2001,JeMoSo1999}); this is the sum of a contribution
due to low-energy virtual photons of
$-27.3$~\cite[Eq.~(5.116)]{Pa2001}, and a relatively
small high-energy term of about
$-3.7$~\cite[Eq.~(6.102)]{Pa2001}. In estimating these contributions,
we follow~\cite{PaJe2003}.

This leads to
the following overall result for the $B_{60}$ coefficients,
where the first two results ($1S$ and $2S$) are quoted from~\cite{PaJe2003},
and the latter results are obtained within the current 
investigation:
\begin{subequations}
\label{b60s}
\begin{eqnarray}
B_{60}(1S) &=& -61.6(3)\pm 15\% \label{b601s}\,, \\[2ex]
B_{60}(2S) &=& -53.2(3)\pm 15\% \label{b602s}\,, \\[2ex]
B_{60}(3S) &=& -51.9(6)\pm 15\% \label{b603s}\,, \\[2ex]
B_{60}(4S) &=& -51.0(8)\pm 15\% \label{b604s}\,, \\[2ex]
B_{60}(5S) &=& -50.3(8)\pm 15\% \label{b605s}\,, \\[2ex]
B_{60}(6S) &=& -49.8(8)\pm 15\% \label{b606s}\,.
\end{eqnarray}
\end{subequations}
The values are in numerical agreement with those used in 
latest adjustment of the fundamental physical 
constants~\cite{MoTaPriv2004}; these are based on an 
extrapolation of the results obtained for 
$n=1,2$~\cite{PaJe2003} to higher $n$, using 
a functional form $a + b/n$, with an extra uncertainty
added in order to account for the somewhat incomplete form 
of the functional form used in the extrapolation.
We here confirm the validity of the approach taken in~\cite{MoTaPriv2004}
by our explicit numerical calculation.

%
%
\section{CONCLUSIONS}
\label{conclu}

The calculation of binding two-loop self-energy corrections has received
considerable attention within the last decade~\cite{Pa1993pra,%
Pa1994prl,EiGrPe1994,EiSh1995}. As outlined in Sec.~\ref{intro},
there is an intuitive physical reason why a reliable understanding of the 
two-loop energy shift requires the 
calculation of all logarithmic as well as nonlogarithmic corrections
through the order of $\alpha^2\,(Z\alpha)^6\,m$.
It is the order of $\alpha^2\,(Z\alpha)^6\,m$ which is the ``natural''
order-of-magnitude for the two-loop self-energy effect from 
the point of view of nonrelativistic quantum electrodynamics (NRQED);
i.e.,~low-energy virtual photons begin to contribute at this order only,
whereas effects of lower order [$\alpha^2\,(Z\alpha)^4\,m$
and $\alpha^2\,(Z\alpha)^5\,m$]
are mediated exclusively by high-energy virtual quanta.

In Sec.~\ref{knowncoeff}, we recall known lower-order  
coefficients for $S$ states, as well as logarithmic 
corrections. The formulation of the problem
within NRQED~\cite{CaLe1986} and the actual numerical
evaluation of the two-loop Bethe logarithms $b_L$ for higher
excited $S$ states is discussed in Sec.~\ref{2Lnrqed}.
Numerical results for $b_L$ are given in Eq.~(\ref{bL}).  
As shown in Fig.~\ref{fig5},
the dependence of these results on the principal quantum number
follows a pattern recently observed quite universally for 
binding corrections to radiative bound-state energy 
shifts~\cite{Je2003jpa,JeEtAl2003}.
This permits an extrapolation of the results to higher 
principal quantum numbers, which is useful for the 
determination of fundamental constants~\cite{MoTaPriv2004}.

There is a certain ambiguity in the definition of the 
two-loop nonlogarithmic coefficient $B_{60}$ due to 
squared decay rates~(Sec.~\ref{ambiguity});
this aspect has previously been considered in~\cite{JeEvKePa2002} 
for $P$ states. Here, the treatment of the squared decay rates is
generalized to excited $S$ states. The ambiguity, while formally
of the order of $\alpha^2\,(Z\alpha)^6\,m$,
is shown to be barely significant for $S$ states (see Table~\ref{table2}), 
due to small prefactors. 

Numerical estimates of the total $B_{60}$-coefficient 
for excited $nS$ states ($n=1,\dots,6$) are given 
in Eq.~(\ref{b60s}). These results improve our theoretical knowledge of
the hydrogen spectrum. On the occasion, we would also 
like to mention ongoing efforts regarding the 
calculation of binding three-loop corrections 
of order $\alpha^3\,(Z\alpha)^5$~\cite{EiSh2004}.
At $Z=1$, these binding three-loop corrections are
of the same order-of-magnitude ($\alpha^8$) as 
the two-loop Bethe logarithms discussed here. There is considerable 
hope that in the near future, 
our possibilities for a self-consistent interpretation
of high-precision laser spectroscopic experiments
may be enhanced significantly via a combination
of ongoing experiments at Paul Scherrer Institute (PSI),
Laboratoire Kastler--Brossel and Max--Planck--Institute for 
Quantum Optics, whose purpose is a much improved Lamb-shift
measurement ($1S$--$2S$-- and $1S$--${3S}$--transitions
combined with an improved knowledge
of the proton charge radius as derived from the 
PSI muonic hydrogen experiment).
The comparison of numerous transitions
in hydrogenlike systems with theory
may also help in this direction as it allows 
for an evaluation of the proton charge radius 
using an overdetermined system of equations,
provided that theoretical Lamb-shift values are used
as input data for the systems of equations rather 
than variables for which the systems should be solved
[see, e.g., Eqs.~(2) and (3) of~\cite{UdEtAl1997}].

Finally, we would like comment on 
the relation of the analytic approach ($Z\alpha$ expansion)
pursued here and numerical calculations
of the self-energy at low nuclear charge $Z$,
which avoid the $Z\alpha$ expansion
and which have been carried out on the one-loop level
for high nuclear charge numbers~$Z$~\cite{Mo1974b},
with recent extensions to the numerically more problematic 
regime of low $Z$~\cite{JeMoSo1999}.
One might note that traditionally, the most accurate
Lamb-shift values at low $Z$ have been obtained via 
a combination of analytic and numerical 
techniques---i.e.,~by 
using both numerical data obtained for high $Z$ 
and known analytic coefficients from the 
$Z\alpha$ expansion~\cite{Mo1975}.
(This is one of the main motivations for pursuing both 
numerical and analytic calculations of the 
two-loop self energy, in addition to the obvious 
requirement for an additional cross-check of the two distinct approaches.)
The general paradigm is the extrapolation of the 
self-energy {\em remainder} function obtained from 
high-$Z$ numerical data after the subtraction
of known analytic terms; in many cases
this extrapolation leads to more accurate 
predictions for the remainder at low $Z$ than 
the simple truncation of the $Z\alpha$ expansion alone. 
Various algorithms have been developed for this 
purpose (see, e.g.,~\cite{Mo1975,IvKa2001proc}).
Indeed, the {\em combination} of analytic and numerical 
techniques has recently proven to be useful in the 
context of binding corrections to the one-loop bound-electron
$g$ factor~\cite{PaJeYe2004}, although direct numerical evaluations
at $Z=1$ had been available before~\cite{YeInSh2002}.
Still, it was possible to improve the theoretical
predictions for the $g$ factor 
self-energy remainder function at low $Z$ by
an order of magnitude via a combination of 
the analytic and numerical approaches, in addition to the 
fact that an important cross-check of 
the analytic and the numerical approaches versus each other became feasible.

\vspace*{0.5cm}

%
%
\acknowledgments

Insightful and elucidating conversations with Krzysztof
Pachucki are gratefully acknowledged.
The author also acknowledges helpful remarks by Peter J. Mohr.
Sabine Jentschura is acknowledged for carefully reading the
manuscript. The stimulating atmosphere at the National Institute
of Standards and Technology has contributed to 
the completion of this project.


\begin{thebibliography}{69}
\bibitem[{rem({\natexlab{b}})}]{remark3S}
\bibinfo{howpublished}{Squared decay rates arise in the order
  $\alpha^2\,(Z\alpha)^6\,m$ only for those states which may decay via the
  electric-dipole channel. This means that the hydrogenic $1S$ and $2S$ states
  remain largely free from any ambiguity. Consequently, we only analyze the
  $3S$, $4S$, $5S$ and $6S$ states in the current section.}
\end{thebibliography}
\end{document}